\documentclass[sigconf,nonacm]{acmart}
\AtBeginDocument{%
  }

\setcopyright{acmcopyright}
\copyrightyear{2023}
\acmYear{2023}
\acmDOI{XXXXXXX.XXXXXXX}

\acmPrice{15.00}
\acmISBN{978-1-4503-XXXX-X/18/06}

\usepackage{xspace}
\usepackage{cleveref}
\usepackage{subcaption}
\usepackage[most]{tcolorbox}
\usepackage{tikz}
\usepackage{multirow}

\newcommand*\circled[1]{\tikz[baseline=(char.base)]{
            \node[shape=circle,draw,inner sep=0.7pt,fill=black,text=white] (char) {\textbf{\texttt{#1}}};}}
\newcommand*\smallcircled[1]{\tikz[baseline=(char.base)]{
            \node[shape=circle,draw,inner sep=0.0pt,fill=black,text=white] (char) {\textbf{\texttt{#1}}};}}
\newcommand*\bluecircled[1]{\tikz[baseline=(char.base)]{
            \node[shape=circle,draw,inner sep=0.7pt,fill=blue,text=white] (char) {\textbf{\texttt{#1}}};}}

\newcommand{\name}{\textsc{AutoSD}\xspace}

\begin{document}

\title{Explainable Automated Debugging via\\Large Language Model-driven Scientific Debugging}

\author{Sungmin Kang}
\authornote{This work was done as part of an internship at Microsoft Research Asia.}
\email{sungmin.kang@kaist.ac.kr}
\affiliation{
    \institution{KAIST}
    \city{Daejeon}
    \country{South Korea}
}

\author{Bei Chen}
\email{bei.chen@microsoft.com}
\affiliation{
    \institution{Microsoft Research Asia}
    \city{Beijing}
    \country{China}
}

\author{Shin Yoo}
\email{shin.yoo@kaist.ac.kr}
\affiliation{
    \institution{KAIST}
    \city{Daejeon}
    \country{South Korea}
}

\author{Jian-Guang Lou}
\email{jlou@microsoft.com}
\affiliation{
    \institution{Microsoft Research Asia}
    \city{Beijing}
    \country{China}
}

\renewcommand{\shortauthors}{Kang et al.}

\begin{abstract}
Automated debugging techniques have the potential to reduce developer effort 
in debugging, and have matured enough to be adopted by industry. However, one 
critical issue with existing techniques is that, while developers want 
rationales for the provided automatic debugging results, existing techniques 
are ill-suited to provide them, as their deduction process differs 
significantly from that of human developers. 
Inspired by the way developers interact with code when debugging, we propose 
Automated Scientific Debugging (\name), a technique that given buggy code and 
a bug-revealing test, prompts large language models to automatically 
generate hypotheses, uses debuggers to actively interact with buggy code, and thus automatically reach conclusions prior to patch generation. By aligning the reasoning of automated debugging more closely with that of human developers, we aim to produce intelligible explanations of how a specific patch has been generated, with the hope that the explanation will lead to more efficient and accurate developer decisions. 
Our empirical analysis on three program repair benchmarks shows that \name performs competitively with other program repair baselines, and that it can indicate when it is confident in its results.
Furthermore, we perform a human study with 20 participants, including six professional developers, to evaluate the utility of explanations from \name. Participants with access to explanations could judge patch correctness in roughly the same time as those without, but their accuracy improved for five out of six real-world bugs studied: 70\% of participants answered that they wanted explanations when using repair tools, while 55\% answered that they were satisfied with the Scientific Debugging presentation.
\end{abstract}




\maketitle

\section{Introduction}

Automated debugging techniques, such as Fault Localization (FL) or Automated Program Repair (APR), aim to help developers by automating the debugging process in part. Due to the significant amount of developer effort that goes into debugging~\cite{zeller2009programs}, automated debugging is a research topic of significant interest~\cite{Liu2020ot}: many papers are published every year~\cite{monperrus2020livingreview}, and the field is mature enough to see adoption by industry~\cite{Kirbas2021BloombergAPR,SapFix2018Marginean}.

Regarding the practical adoption of these techniques, a body of literature surveying developer expectations on automated debugging has consistently highlighted that, as much as strong performance on software engineering tasks is important, so is supporting information that helps developers judge the results. For example, Kochhar et al.~\cite{Kochhar2016FLExpect} perform a study of developer expectations on fault localization, and find that more than 85\% of developers agree that the ability to provide rationale is important. Further, Kirbas et al.~\cite{Kirbas2021BloombergAPR} note that some developers responded negatively to automated program repair results, citing that they would come ``out of the blue''. Such findings suggest that strong automated debugging results may not be acceptable on their own, and may need supporting information that helps \emph{explain} the results.

Despite the consistent request for explainable processes for automated results, to the best of our knowledge explainable automated debugging techniques can be difficult to come by. For example, in the living review of APR compiled by Monperrus updated in August 2022~\cite{monperrus2020livingreview}, the word `explain' appears only in one position paper~\cite{Monperrus2019ExpAPRVision}, revealing that the critical research on how to explain repair suggestions to developers is under-explored. We argue that this is in part because existing automated debugging techniques reason in starkly different ways to humans. Whereas existing automated debugging techniques will reduce a search space~\cite{Jiang2018ShapingPR} and try multiple solutions to find results that are correlated with the location and fix of a bug~\cite{MUSE2014Moon}, human developers will generally utilize debuggers and \texttt{print} statements to interact with the buggy code, understand its behavior and in turn make a patch based on such observations~\cite{DevDebug2014Siegmund}. That is, the reasoning traces~\cite{WNWN2009Lim} of existing automated debugging processes are so different from those of developers, that suggesting them may contribute little to the understanding of a generated patch.

As a step towards automated debugging techniques that can generate explanations that help developers, we propose \name, which bridges the gap between humans and automated debugging processes. To do so, \name leverages Large Language Models (LLMs) and a debugger interface to automatically emulate the Scientific Debugging (SD) process for developers proposed by Zeller~\cite{zeller2009programs}. \name prompts an LLM to automatically generate hypotheses about what is causing the bug, along with a debugger script that would test the hypotheses. \name then executes the suggested debugger command and provides the LLM with the result; based on this, the LLM finally decides whether the hypothesis was met, and predicts if the debugging process is done, or additional investigation is required. The intermediate debugging text generated as a result can naturally be presented as an \emph{explanation} describing how \name reached its conclusion. Emulating Scientific Debugging has ideal properties for explainable debugging: notably, as existing work identifies that developers use the principles of Scientific Debugging to debug even without formal training~\cite{DevDebug2014Siegmund}, the explanations could help inform or augment the thought process of developers.

We empirically evaluate \name by first evaluating it on three program repair benchmarks. Our results indicate \name can achieve competitive repair results to non-explainable APR techniques. In terms of practical usage, precision is an important factor~\cite{Xiong2017ACS}; we find that for cases when \name indicates it had collected enough information for debugging, repair performance is in fact higher. As language models become more capable, the repair performance of \name rapidly increases as well, demonstrating the potential of \name.
We further perform a user study on Python developers involving 20 participants, including six professional developers, under a realistic APR application setting: reviewing patches for acceptance. Our results demonstrate that the debugging traces generated by \name enhance developer accuracy in terms of accessing whether the patch is correct for 83\% of the real-world bugs studied, while keeping the amount of time in which developers could judge whether the patch roughly constant; these results suggest that humans benefit from the automatically generated patch explanations of \name. Furthermore, 70\% of participants responded that they would see explanations as an important factor when using APR tools, and 55\% were satisfied with the Scientific Debugging formulation of \name.

Overall, our contribution may be summarized as:

\begin{itemize}
  \item We identify that explainable automated debugging may be achieved by LLMs emulating developer processes, and as a demonstration propose \name, which uses LLMs to emulate Scientific Debugging~\cite{zeller2009programs};
  \item We perform empirical analyses on three APR benchmarks, demonstrating that \name can achieve significant APR performance while also generating explanations;
  \item We conduct a developer study on \name, based on a realistic scenario of patch review, and demonstrate explanations from \name can aid developers in decision-making;
  \item We further solicit feedback from users regarding repair explanations, presenting a guideline for future improvement of explanations.
\end{itemize}

The remainder of the paper is organized as follows. We introduce the technical background to our work in \Cref{sec:background}, and our technique \name in \Cref{sec:autosd}. The evaluation setup and research questions are provided in \Cref{sec:eval_setup}, and the empirical results based on these experiments are presented in \Cref{sec:results}. Threats and limitations are discussed in \Cref{sec:discussion}, and \Cref{sec:conclusion} concludes.

\section{Background}
\label{sec:background}
This section provides the motivation and background for our work.

\subsection{Explainable Automated Debugging}
\label{sec:motivation}

Automated debugging has a long history, with research often being done on the topics of fault localization~\cite{MUSE2014Moon,Jones2002Tarantula,Xia2019DeepFL} and automated program repair~\cite{Gazzola2019APRSurvey}. As described before, while the technical complexity and performance of automated debugging techniques has been increasing~\cite{jiang2023knod}, including the use of LLMs for APR~\cite{jiang2023impact,xia2022practical}, empirical work on explaining results for developer consumption has been difficult to identify. In addition to Monperrus' living review on APR having only one paper mentioning explanations~\cite{monperrus2020livingreview}, Winter et al.~\cite{Winter2022DevAPR} find 17 human studies evaluating APR, of which none involved explanations directly from an APR tool; Kochhar et al.~\cite{Kochhar2016FLExpect} survey fault localization techniques at the time, and find two techniques that could provide explanations of their results~\cite{Sun2013MiningSig,Mariani2011DiagnoseFaults}; unfortunately, both papers did not have human studies. 

This contrasts to the growing body of literature showing that, to adopt automated debugging techniques in practice, `explanations' for the results would be welcome. Developers have stated their desire for explanations in multiple occasions: along with the findings of Kochhar et al.~\cite{Kochhar2016FLExpect} mentioned earlier, a developer study on expectations for APR by Noller et al.~\cite{noller2022APRTrust} notes that ``the most commonly mentioned helpful output from an APR tool is an \textit{explanation} ... including its \textit{root cause}''. Developer expectation is particularly important because when automated debugging has been adopted by industry, automatically generated patches are consistently reviewed by developers. At Meta, the APR system is connected to the internal code review platform~\cite{SapFix2018Marginean}; at Bloomberg, Kirbas et al.~\cite{Kirbas2021BloombergAPR} write that ``Bloomberg's view was that full automation was far from ideal'', and they subject APR patches to be reviewed by a software engineer. This is also reflected in Noller et al.'s results that ``full developer trust requires a manual patch review''. 

A promising way to present developers with explanations could be to show the reasoning trace~\cite{WNWN2009Lim} of a tool, i.e. how an automated debugging tool came to recommend a certain line for FL or a certain patch for APR. Unlike post-hoc explanation techniques such as commit message generation~\cite{CommitGen2017Jiang}, such reasoning traces can answer critical questions that a developer may have, such as `why this patch?'; indeed, research in Human-Computer Interactions (HCI) have indicated that explanations should strive to be capable of answering \emph{why} an approach gave a certain result~\cite{WNWN2009Lim}.

However, current automated debugging techniques are ill-suited to generate helpful explanations for their results, as their reasoning traces deviate from human reasoning traces significantly. Using a common classification of APR techniques~\cite{Goues2019APR} as an example\footnote{The explainability of FL is discussed in the appendix.}, generate-and-validate (G\&V) techniques~\cite{Gazzola2019APRSurvey} (which includes learning-based techniques~\cite{Jiang2018ShapingPR,Xia2022Alpharepair,Zhu2021Recoder}) will generate variants of the buggy code until a test passes. As their deduction process is simply enumerating changes and trying them one by one, the process runs without regard to any `root cause'. Semantics-based APR techniques such as Angelix~\cite{Angelix2016Mechtaev} use variable values as inputs to Satisfiability Modulo Theory (SMT) solvers to more effectively search within a patch space; thus they are not inherently identifying any `root cause' either. This is not to say these techniques are ineffective at fixing bugs - numerous work on APR shows that existing APR techniques can fix a wide array of bugs. Rather, we argue that because their reasoning trace is so different from humans, it is difficult to make a satisfactory explanation of their results. On the other hand, one way to make satisfactory explanations would be to develop an automated debugging technique that deduces in a similar way to humans, to make the decision-making process transparent~\cite{Dam2018ExplainableSA}.

\subsection{Scientific Debugging}
\label{sec:zellersd}

To align APR reasoning traces more closely to those of human developers, we must know how developers debug in practice. Previous work on developer debugging patterns provide glimpses into how debugging is actually done.

Early work on developer debugging found that there was a ``gross descriptive model'' that developers followed, in which developers formulated hypotheses, then verified whether the hypotheses are true~\cite{DevDebugOld1975Gould}. A formal version of this process was named \emph{Scientific Debugging} by Zeller~\cite{zeller2009programs}, who advocated for developers to maintain a debugging log consisting of an iteration of the following items:

\begin{itemize}
    \item Hypothesis: a tentative description that explains the bug and is consistent with the known observations;
    \item Prediction: an expected outcome if the hypothesis is true;
    \item Experiment: a means of verifying the prediction;
    \item Observation: the result of an experiment;
    \item Conclusion: a judgement of the hypothesis, based on the observation.
\end{itemize}

Siegmund et al.~\cite{DevDebug2014Siegmund} found that even without formal training in debugging techniques, all developers surveyed would roughly follow the `hypothesis formulation, then verification' process of scientific debugging. Thus, Scientific Debugging can be seen as a formal way of describing the dominant developer thought process when debugging, and thus we seek to emulate this process to make an explanation when generating APR results.

\subsection{Large Language Models}
\label{sec:promptE}

In this paper, we seek to emulate the Scientific Debugging process via Large Language Models (LLMs). We believe LLMs are capable of emulating Scientific Debugging for the following reasons. First, they have shown increasingly strong performance on question-answering benchmarks that involve reasoning~\cite{brown2020language,openai2023gpt4}, which also makes it possible that they would be capable of predicting whether a hypothesis is met, and which hypothesis to investigate next. While it would be difficult to manually gather a large amount of data that contains debugging traces in the Scientific Debugging format, LLMs have also been demonstrated to be capable of few-shot or zero-shot problem solving: that is, given a few examples or simply a description of the task to be solved in the form of a natural-language \emph{prompt}, they are capable of doing the task~\cite{brown2020language}. This capability improves with Reinforcement Learning with Human Feedback (RLHF) training~\cite{ouyang2022training}, which the main LLM of our task (ChatGPT of OpenAI) was trained on. Finally, the interaction with code that Scientific Debugging asks for requires the use of external tools. When using `Chain-of-Thought' (CoT) prompting~\cite{Wei2022ChainOT}, LLMs appear capable of using the results of external tools to improve their performance as well~\cite{yao2022react, gao2022pal}. As a result, we believe that LLMs are well-positioned to emulate the Scientific Debugging process, and thus generate reasoning traces intelligible to developers.

\begin{figure*}
    \includegraphics[width=\textwidth]{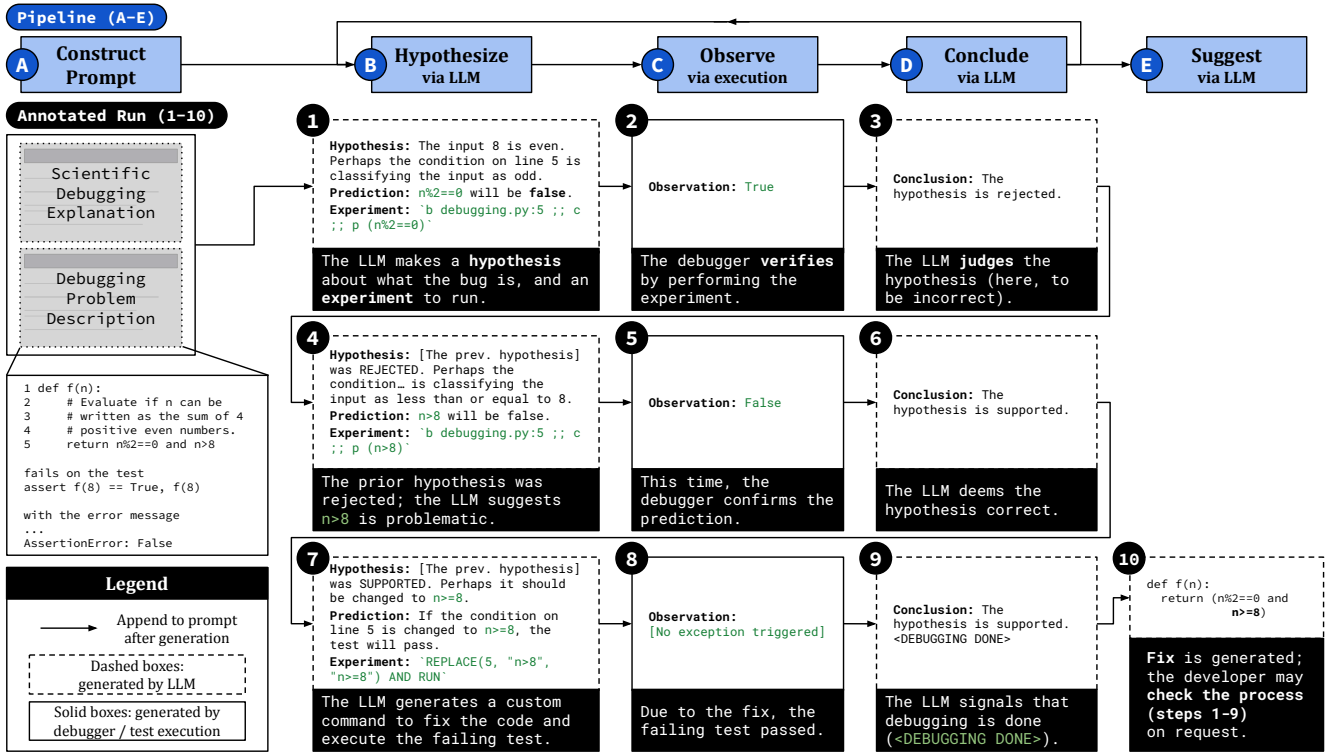}
    \caption{The pipeline and a real example run of \name, with annotations in black boxes and lightly edited for clarity. Given a detailed description of the scientific debugging concept and a description of the bug (A), \name will generate a hypothesis about what the bug is and construct an experiment to verify, using an LLM (B), actually run the experiment using a debugger or code execution (C), and decide whether the hypothesis is correct based on the experiment result using an LLM (D). The hypothesize-observe-conclude loop is repeated until the LLM concludes the debugging or an iteration limit is reached; finally, a fix is generated (E), with an \emph{explanation} (white boxes from (1) to (9)) that the developer may view.}
    \label{fig:teaser}
\end{figure*}

\section{Automated Scientific Debugging}
\label{sec:autosd}

The overall process of our approach is presented in \Cref{fig:teaser}. To start, the prompt containing relevant information is generated (\Cref{fig:teaser} \bluecircled{A}): this consists of a detailed explanation of what Scientific Debugging is, and a description of the debugging problem itself, so that \name can proceed with the following steps. With the initial prompt prepared, \name generates a hypothesis on what is wrong with the code or how it can be fixed, along with the concrete experiment that would validate such a hypothesis, using an LLM (\Cref{fig:teaser} \bluecircled{B}). The experiment script will be passed to a background debugger/code executor process, which runs the script and returns the actual result (\Cref{fig:teaser} \bluecircled{C}). Based on the observed information, \name decides whether the hypothesis was verified or not using an LLM (\Cref{fig:teaser} \bluecircled{D}); depending on the conclusion, \name either starts with a new hypothesis or opts to terminate the debugging process and generate a fix. When the interaction with the code is over, \name generates a bug fix based on the gathered information (\Cref{fig:teaser} \bluecircled{E}). Unlike other automated program repair techniques we are aware of, as a result of steps (\bluecircled{B} - \bluecircled{D}) \name can provide a \emph{rationale} of how a particular fix was generated, which can then be provided to the developer upon request.

\subsection{Constructing the Input Prompt}

To construct the initial prompt, as in the example presented in \Cref{fig:teaser} \bluecircled{A}, we first manually wrote a detailed description of Scientific Debugging that explains what hypotheses, predictions, experiments, observations, and conclusions are, along with multiple examples for each category, so that the LLM can generate an intelligible reasoning trace. The full description can be found in the appendix; here, we describe the aspects of the description critical for the pipeline of \name in detail. For one, concrete examples of experiments are provided, to allow the LLM to predict appropriate experiment scripts: composite debugger commands (consisting of setting a breakpoint, running code, and printing a value) and a Domain-Specific Language (DSL) that we define to allow edit-and-execute commands are given. Regarding the DSL, the prompt specifies that the following commands are available: \texttt{REPLACE(line, old\_expr, new\_expr)} that changes an expression at \texttt{line}, \texttt{ADD(line, new\_expr)} that adds a new statement above \texttt{line}, and \texttt{DEL(line, old\_expr)} that allows deletion of an expression in a line. Multiple commands can be joined with the \texttt{AND} connector, and finally the bug-revealing test can be executed after modification via the \texttt{RUN} command. In addition to experiment commands, the prompt instructs to predict the \texttt{<DEBUGGING DONE>} token (\texttt{<DONE>} for short in the rest of the paper) if enough information to discern the patch has been gathered, so that we can gauge how confident \name is in its patch. The prompt is detailed enough so that our default LLM, ChatGPT, can follow the instructions zero-shot, i.e., without a concrete demonstration of the full process.  On this description of scientific debugging, we add the bug-specific information: concretely, the buggy function/method, the test that reveals the bug, the error message when the bug is executed, and if available a bug report. We add this information as we believe such information would be necessary, if not sufficient, for a human to debug an issue, and thus would likely also help an automated technique to predict appropriate hypotheses and ultimately succeed in debugging.

\subsection{Hypothesize-Observe-Conclude}

With the initial prompt, \name starts iterating over the `hypothesize-observe-conclude' loop depicted in \Cref{fig:teaser} (\bluecircled{B} - \bluecircled{D}). The result of each process is appended to the prompt to allow incremental hypothesis prediction; i.e. when generating the conclusion in \circled{3}, the LLM would predict it based on the concatenation of the initial prompt, \circled{1}, and \circled{2}. We describe each iteration of the loop as a \emph{step}: for example, \Cref{fig:teaser} \circled{1} - \circled{3} would make up one step.

\textbf{Hypothesize.} Here, we lead the language model to generate a hypothesis by appending the token \texttt{Hypothesis:} to the prompt, so that the language model generates a hypothesis about the bug. We observe that the \texttt{Prediction:} and \texttt{Experiment:} line headers are also generated in turn by the LLM, due to the detailed description of the scientific debugging process provided by the prompt. The important aspect for the next step is the \texttt{Experiment} command, where the language model either generates a debugger command that can be executed by a debugger, or a custom code modification-and-execution script so that the language model can `test' a certain change. As the document is in Markdown format, the \texttt{Experiment} script is wrapped in backticks (\texttt{\textasciigrave}); this script is extracted from the LLM output to get concrete code execution results in the next step. Examples can be seen in \Cref{fig:teaser} \circled{1}, \circled{4}, and \circled{7} - note that \name also localizes the fault as a part of the hypothesizing process, thus making fault localization explainable as well.

\textbf{Observe.} The generated experiment script is passed to a background process based on traditional software engineering tools that provides real execution results back to the language model, so that we can ground the generation process of \name on real results, and also build credibility for developer presentation. The model can either (i) invoke a composite debugger command by setting a breakpoint and printing a value, or (ii) modify the code and run the failing test with the aforementioned DSL. When executing a debugger command, it is executed via the command-line interface of the language-appropriate debugger, and the output from the last subcommand of the composite command (assumed to be a \texttt{print} command) is returned, as in \Cref{fig:teaser} \circled{2} and \circled{5}. When the breakpoint is within a loop, the debugger collects values at different timesteps of execution and returns them together, e.g. `At each loop execution, the expression was: [v1, v2, ...]', up to a maximum of 100 values. Meanwhile, upon test execution from a edit-and-execute DSL command, if an exception is raised, the exception type and message are returned as the observation; otherwise, the result `[No exception triggered]' is appended, as in \Cref{fig:teaser} \circled{8}.

\textbf{Conclude.} Based on the observation, \name invokes the LLM to check whether the hypothesis and the observation are consistent, by having the LLM predict if the hypothesis is rejected (e.g. \circled{3}), supported (e.g. \circled{6}), or undecided due to an unexpected observation. We have the LLM generate the conclusion to maximize flexibility in value interpretation. As described earlier, the LLM may predict a separate \texttt{<DONE>} token at this step if it predicts the debugging process is complete; in such cases, \name would have greater confidence in its output. An example is shown in \Cref{fig:teaser} \circled{9}: on the information that the previously failing test now passes, the LLM concludes that debugging is done. If the \texttt{<DONE>} token is predicted, \name proceeds to generate a fix as in \Cref{sec:fixsugg}; otherwise the loop restarts with hypothesizing based on the newly available information until a maximum iteration limit $s$ is reached. If \texttt{<DONE>} is not predicted until then, \name is failing to identify the cause of the bug, and we may be more skeptical of the generated patch.

\subsection{Fix Suggestion}
\label{sec:fixsugg}
When \name has completed its interaction with the code, the conclusions to each of the hypotheses are assessed, and rejected hypotheses are automatically removed from the prompt prior to patch generation, as this empirically improved program repair performance in our experiments. Even if rejected hypotheses are not involved when making the fix itself, rejected hypotheses can still be presented to the developer as context for successful hypotheses. We subsequently prompt the LLM to generate a fix using the available information by appending the words ``\texttt{The repaired code (full method, without comments) is:\textbackslash{}n\`{}\`{}\`{}}''. This prompt leads the LLM to generate repaired code, based on the information available from the problem description and the code interaction, as in \Cref{fig:teaser} \smallcircled{10}. Identically to other APR techniques, a patch is ultimately generated; what makes \name unique is that it can show its \emph{intermediate reasoning steps} (\circled{1} - \circled{9}) as an \emph{explanation} that can help the developer understand where a patch comes from.

\section{Evaluation Setup}
\label{sec:eval_setup}

Here we describe the setup for our empirical evaluation.

\subsection{Research Questions}
\textbf{RQ1: Feasibility.} While the main focus of our work is to generate a reasoning chain for automated debugging results, good performance in the debugging task itself is also important~\cite{Kochhar2016FLExpect,noller2022APRTrust}. We thus seek to answer whether \name achieves performance competitive to prior APR techniques, and when compared to prompting an LLM to immediately predict a fix (this baseline is referred to as \textsc{LLM-Base} in the rest of the paper). We aim to demonstrate that the explainability of \name does not come with a significant performance cost, even as prior reviews on explainable AI describe a tradeoff between interpretability and performance~\cite{arrieta2020explainable}. We evaluate \name on the Almost-Right HumanEval benchmark we construct to mitigate data leakage concerns, and the Defects4J v1.2 and 2.0 benchmarks \cite{Rene2014dj} consisting of real-world bugs.

\textbf{RQ2: Debugger Ablation.} In this research question, we first evaluate whether the performance of \name is better when it indicates that debugging is done via the \texttt{<DONE>} token; as precision is important for practical tools for developers, if \name can indicate when it is likely to be correct, this would aid developer adoption of \name. Based on our confidence-with-\texttt{<DONE>} experiments, we evaluate the performance of \name when debuggers are not used, and observations are `hallucinated' by the LLM instead of obtained via actual code execution. We evaluate whether under this setting, the \texttt{<DONE>} token continues to be a marker of strong performance. 

\textbf{RQ3: Varying LLM.} We evaluate the performance of \name as we vary the LLM that is used. While we empirically found the best performance when using the ChatGPT model, and thus used it as the default setting throughout the rest of the paper, by varying the size of the language model and plotting the performance, we investigate automated repair performance as models improve in terms of parameter size and training sophistication.

\textbf{RQ4: Developer Benefit.} Via our human study, we evaluate whether developers benefit materially from automatically generated explanations by \name, i.e. regardless of their opinion towards explanations. In our human study, participants are given the buggy code, a bug-revealing test, a candidate patch, and half of the time an explanation, and asked to determine whether the patch correctly addresses the issue that the test reveals. We measure the time and accuracy of developers when deciding whether a patch is correct, along with developer answers to the question `did the explanation help you make the decision?'. We thus hope to evaluate whether developers benefit by being provided explanations.

\textbf{RQ5: Developer Acceptance.} We evaluate whether the explanations of \name are acceptable to developers by asking them six questions on whether they would want to use APR, whether they would want explanations when using APR, and whether \name and each element of its explanation were satisfactory. We thus hope to measure whether developers are willing to use explanations, distinctly from whether their productivity increases from explanations. We additionally perform interviews to identify what developers liked about the explanations of \name, and what could improve.

\textbf{RQ6: Qualitative Analysis.} We provide examples of liked and disliked patch attempts and their corresponding explanations in this research question as further context, along with a breakdown of common failure causes by analyzing a random sample of 25 cases in which all hypotheses generated by \name were classified as incorrect by itself.

\subsection{Environment}

\subsubsection{Evaluating APR Performance}

To evaluate the performance of \name, we use four program repair benchmarks. First, the widely-used Defects4J benchmarks~\cite{Rene2014dj} version 1.2 and 2.0, which have been used by prior work as a standard benchmark to compare APR techniques~\cite{Liu2020ot}, are used. We also use the BugsInPy benchmark~\cite{Widyasari2020BIP} (abbreviated to BIP in our paper) for the sake of getting real-world Python bugs to evaluate in our human study, but we do not report the performance of \name on BIP as many of its bugs needed additional environment setup not described in the README.

We additionally construct the Almost-Right HumanEval (ARHE) dataset based on the HumanEval Python single-function synthesis benchmark by Chen et al.~\cite{chen2021evaluating}. We do so in the hopes that it will be free from data contamination concerns, as HumanEval was explicitly made by Chen et al. to avoid data contamination when evaluating their LLM, and was also used to evaluate the recent GPT-4 model~\cite{openai2023gpt4}. The ARHE dataset was built by mutating the human solutions in the HumanEval benchmark so that exactly one test fails, making bugs that cause the code to be `almost' right. We end up with 200 bugs to evaluate with using seven mutators; the detailed composition of the dataset by mutator used is provided in the appendix. For comparison, we additionally compare against a template-based APR baseline that has the reverse mutators of those used to construct the dataset, and randomly applies them to the buggy code. We run this baseline 100 times as it is stochastic. Note that 90 bugs of ARHE are created by deletion or string mutation, and consequently are not reversible by the baseline: all the remaining mutations are reversible and therefore can be fixed by our template-based baseline given sufficient time.

Regarding specific APR parameters, for each dataset we provide \name with the buggy method and generate 10 patches, to match the settings in the large-scale empirical work by Jiang et al.~\cite{jiang2023impact}, who evaluate the repair performance of multiple large language models and more traditional learning-based APR techniques. We note our setting assumes less exact information and is thus more difficult: Jiang et al. evaluate with perfect statement-level FL, whereas \name uses perfect method-level FL and the bug report. When evaluating the generated patches, we run the tests provided by each dataset for each bug; a fix that makes all tests pass is deemed a \textit{plausible} patch, and plausible patches are manually inspected to see if they are semantically equivalent with the developer patch. Semantically equivalent fixes are deemed \textit{correct}.

\name requires the use of an LLM and a debugger. For the LLMs, we experiment with the CodeGen~\cite{Nijkamp2022CG}, Codex~\cite{chen2021evaluating} (code-davinci-002), and ChatGPT (a sibling model to InstructGPT~\cite{ouyang2022training}) LLMs, with the ChatGPT LLM being the default model. Different debuggers are used depending on the target language; we use the \texttt{jdb} tool for the Java benchmarks (Defects4J v1.2 and v2.0) and the \texttt{pdb} tool for the Python benchmarks (ARHE and BugsInPy). The maximum iteration limit, $s$, is set to 3.

\subsubsection{Human Study Parameters}

To approximate the real-world impact of \name, we perform a human study by asking participants to review patches, based on the real-world applications of APR~\cite{SapFix2018Marginean,Kirbas2021BloombergAPR}. We specifically sampled 12 bugs where \name made a patch that caused the initially failing test to pass: a random sample of six such bugs from the ARHE dataset (which had complete documentation), and six real-world bugs from the BugsInPy Python dataset~\cite{Widyasari2020BIP}. In our preliminary studies, we found that reviewing 12 patches could take a long time, so we divided the 12 bugs into two groups of six (each containing three ARHE and three BugsInPy bugs) and randomly assigned participants to solve code review problems from one of the groups. A scheme of the code review screen that was presented to participants is shown in \Cref{fig:hs_screen}; a screenshot of the the survey website can be found in the appendix. Our human study received IRB review exemption (IRB-23-054).

For each code review problem, participants are provided with the buggy code, the bug-revealing (failing) test, along with the patch; they are provided with the explanation in a randomly selected three of the six cases. Each step of the explanation has a header, which is a summary of the hypothesis explaining the bug; the header is color-coded based on the predicted conclusion, with supported/rejected/undecided hypotheses being green/red/yellow, respectively, as in \Cref{fig:hs_screen}. Each header can be clicked to reveal the full reasoning process of \name as depicted in \Cref{fig:teaser}. Participants are asked three questions for each patch: (Q1) whether the patch is a correct patch, where they may answer yes, no, or unsure (as a proxy for checking correctness during the code review process~\cite{Sadowski2018GoogleCR}); (Q2) a short justification of their decision in Q1, to filter potential bad-faith answers; and (Q3) when an explanation is available, whether the explanation was helpful in making their decision, to measure the differing impact of explanations for different patches.

\begin{figure}[h!]
    \centering
    \includegraphics[width=1.0\linewidth]{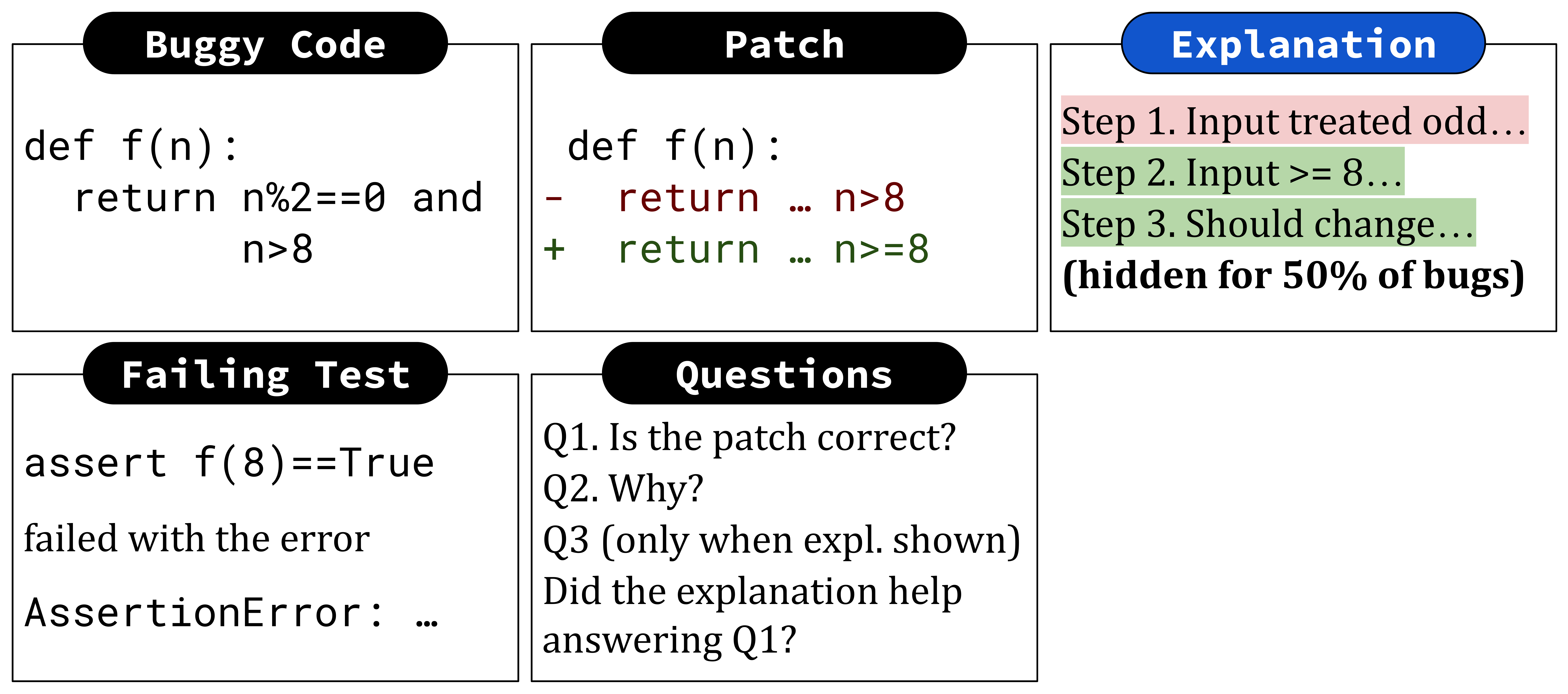}
    \caption{Human Study Screen Scheme} % I know the figure is hideous, it's just a placeholder for now, and I'm not sure if I'll even keep it.
    \label{fig:hs_screen}
\end{figure}

To recruit participants, we advertised the task to both undergraduate and 
graduate students with at least 1 year of Python experience, as well as professional 
developers at a company that specializes in software testing techniques. 
Overall, we recruit 20 participants: eight undergraduate and six 
graduate students, as well as six professional developers whose career span from 3 to 10 years.
Participants start with a briefing of what they should do in the study, solve an example code review problem as practice, and then solve six code review problems in 30-40 minutes in a randomized order. The six code review tasks contain 2 correct and 1 incorrect patches for ARHE and BugsInPy benchmarks, respectively. After conducting a post-questionnaire about their demographics and overall satisfaction with explanations, we perform an interview that lasted about 5 minutes on their impression of the tool for qualitative analysis.

\section{Experimental Results}
\label{sec:results}

We present the results of empirical evaluation below.

\begin{table}[ht]
    \scalebox{0.9}{
    \begin{tabular}{llll}
    \toprule
    \textbf{Result} & \textbf{Template-based} & \textbf{\textsc{LLM-Base}} & \textbf{\name} \\ \midrule
    Plausible     & 85.77 $\pm$ 4.20 & 179 & 189 \\ 
    Correct       &  - & 177 & 187 \\ 
    \bottomrule
    \end{tabular}
    }
    \caption{Repair results on the ARHE benchmark. The template-based performance is based on 100 reruns, and shows the mean and standard deviation repair performance. \label{tab:arhe_results}}
\end{table}

\vspace{-2em}

\subsection{RQ1: Feasibility}
\label{sec:rq1_res}

In \Cref{tab:arhe_results}, we present the APR performance of \name on the ARHE benchmark when compared with \textsc{LLM-Base} and the template-based baseline. Note that the template-based baseline shows significantly weaker repair performance than both \textsc{LLM-Base} and \name when evaluated under the same conditions; as a result, we did not assess correctness for the thousands of patches generated, as the upper bound of correctness is the plausible patch count. Additionally, the performance of \textsc{LLM-Base} and \name are similar, demonstrating \name retains the repair performance of the LLM while simultaneously being capable of generating explanations.

\begin{table}[ht]
    \scalebox{0.9}{
    \begin{tabular}{lllll}
    \toprule
    \textbf{Benchmark} & \textbf{Recoder} & \textbf{InCoder} & \textbf{\textsc{LLM-Base}} & \textbf{\name} \\ \midrule
    D4J v1.2   & 24 & 41 & 87 & 76 \\ 
    D4J v2.0   & 11 & 28 & 110 & 113 \\ 
    \bottomrule
    \end{tabular}
    }
    \caption{Correct repair results on the Defects4J benchmarks. Results for Recoder and InCoder are from Jiang et al.~\cite{jiang2023impact}.~\label{tab:d4j_results}}
\end{table}

In \Cref{tab:d4j_results}, we present the APR performance of \name on the Defects4J benchmarks when compared against \textsc{LLM-Base} and the best-performing techniques from the empirical study by Jiang et al.~\cite{jiang2023impact}: Recoder, a DL-based APR technique~\cite{Zhu2021Recoder}, and finetuned InCoder~\cite{fried2022incoder}, a language model from Facebook, which was finetuned with perfect statement-level FL results, and thus uses more exact information than \name. We find that \name again shows competitive performance when compared to other baselines, even those that have more specific information provided. As an additional reference point, when compared against the repair results of Codex on Defects4J presented by Xia et al.~\cite{xia2022practical} we find that \name outperforms Codex using 200 patch candidates (unlike our 10) on both benchmarks under the `patch function' setting of that paper, which assumes the same FL conditions as our setup. % last sentence is unsatisfactory...

\begin{tcolorbox}[boxrule=0pt,frame hidden,sharp corners,enhanced,borderline north={1pt}{0pt}{black},borderline south={1pt}{0pt}{black},boxsep=2pt,left=2pt,right=2pt,top=2.5pt,bottom=2pt]
    \textbf{Answer to RQ1:} \name is capable of operating at a competitive level of program repair performance when compared to a diverse set of baselines on three repair benchmarks.
\end{tcolorbox}

\begin{figure}[h!]
\centering
\begin{minipage}[t]{.5\linewidth}
  \centering
  \captionsetup{width=.9\linewidth}
  \includegraphics[width=0.9\textwidth]{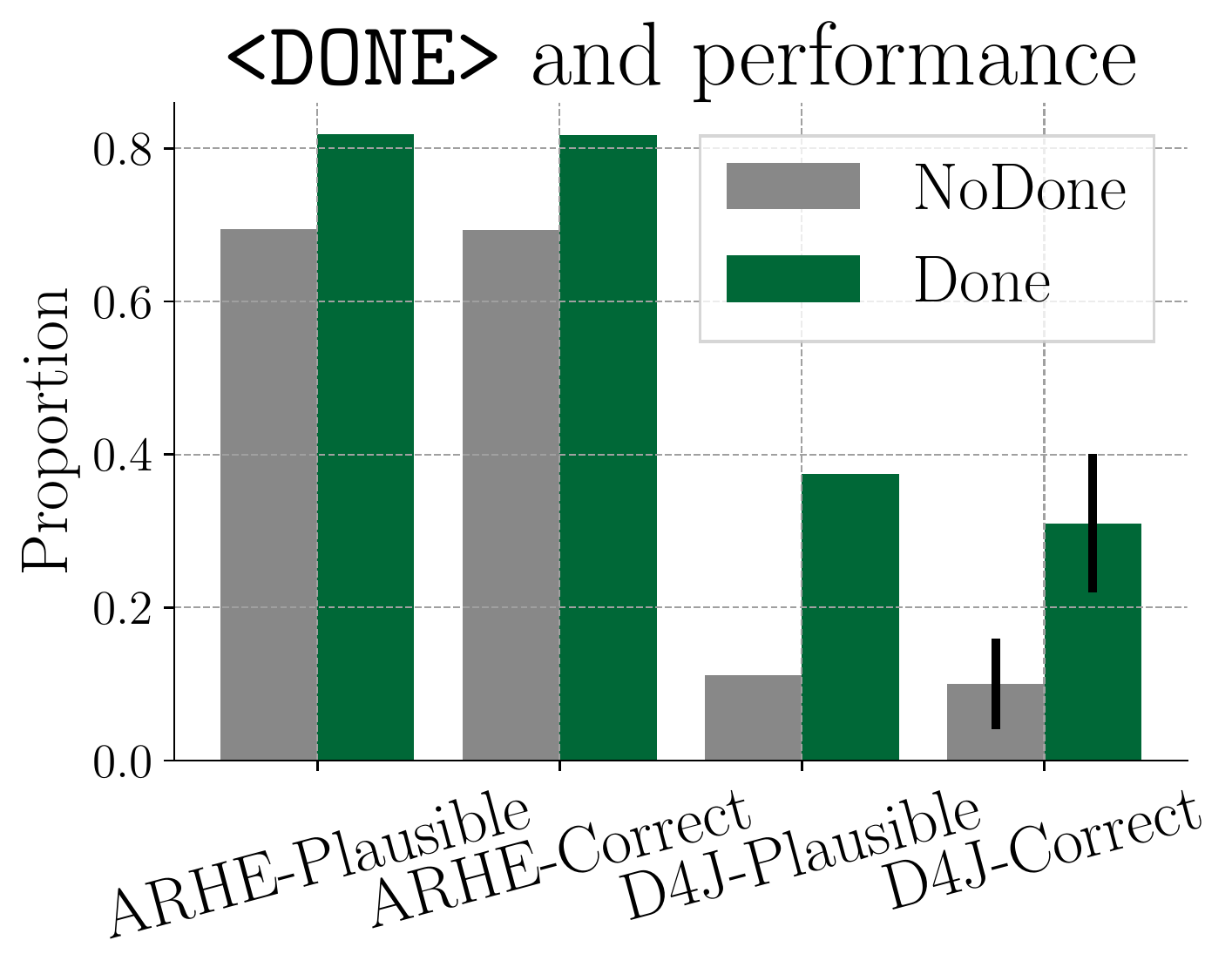}
  \caption{\texttt{<DONE>} \& perf.}
  \label{fig:done_performance}
\end{minipage}%
\begin{minipage}[t]{.5\linewidth}
  \centering
  \captionsetup{width=.9\linewidth}
  \includegraphics[width=0.9\linewidth]{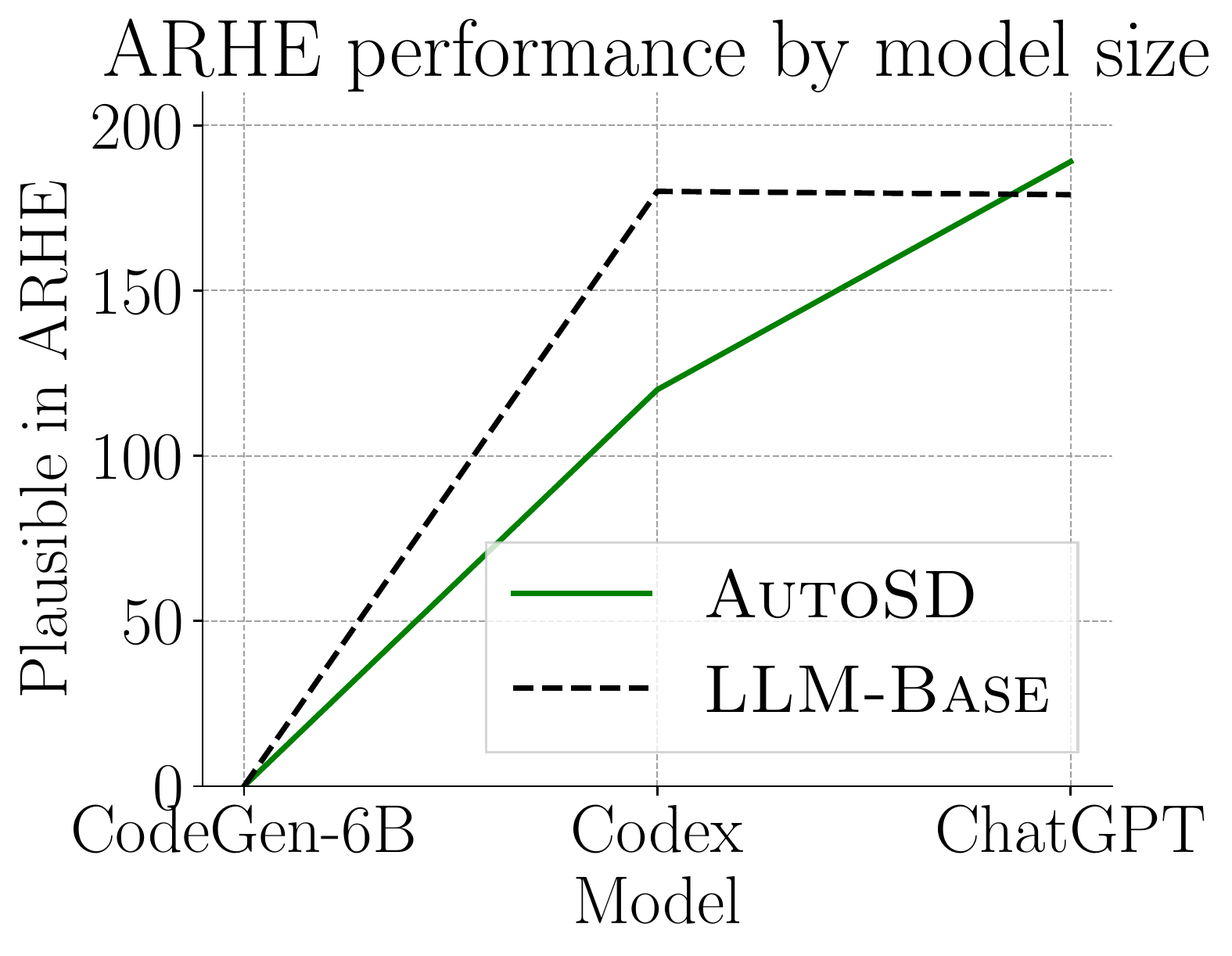}
  \caption{Model Size}
  \label{fig:llm_performance}
\end{minipage}
\end{figure}

\subsection{RQ2: Debugger Ablation}
\label{sec:rq2_res}

This RQ first investigates whether the confidence in a result indicated by the prediction of the \texttt{<DONE>} token actually correlates with better performance. The results are presented in \Cref{fig:done_performance}. For Defects4J, as it was infeasible to manually label all 1045 plausible patches generated for the dataset, we sampled 100 patches with and without \texttt{<DONE>} to get results. As the figure shows, for both the ARHE and Defects4J datasets, \name shows a higher precision when the \texttt{<DONE>} token is generated as part of a conclusion, indicating that \name can indeed signal when it is likely to generate a plausible or correct patch. Furthermore, for bugs where a plausible patch was generated and the \texttt{<DONE>} token was predicted, 89\% were correctly fixed, while for bugs with plausible patches but without \texttt{<DONE>} predictions 82\% were correctly fixed.

We also investigate the performance when the debugger/code execution results are also predicted by the LLM, instead of being obtained via concrete execution, for the ARHE dataset; would the \texttt{<DONE>} token still predict good performance? In this `debugger hallucination' scenario, \texttt{<DONE>}-predicted solutions were actually 11\%p \emph{less} likely to be plausible; this is in contrast to using actual code execution results, where \texttt{<DONE>}-predicted solutions are 12.4\%p more likely to be plausible. Furthermore, individual runs became much less likely to be plausible: while 73\% of the individual \name runs would yield a plausible patch, only 63\% would when the debugger was ablated. Thus, incorporating code execution contributes to the reliability of \name; we later demonstrate in RQ5 that developers found real code execution results useful as well.

\begin{tcolorbox}[boxrule=0pt,frame hidden,sharp corners,enhanced,borderline north={1pt}{0pt}{black},borderline south={1pt}{0pt}{black},boxsep=2pt,left=2pt,right=2pt,top=2.5pt,bottom=2pt]
    \textbf{Answer to RQ2:} \name can indicate when its answers are more likely to be correct with the \texttt{<DONE>} token, which we also use to verify the utility of debugger use.
\end{tcolorbox}

\subsection{RQ3: Varying LLM}
\label{sec:rq3_res}

In \Cref{fig:llm_performance}, we depict the performance of \name as different underlying LLMs are used, with the $x$ axis showing different LLMs roughly sorted in terms of number of parameters and the technical advancement of training, and the $y$ axis showing the performance of \name when using the LLM on the ARHE benchmark. The performance of \name is depicted along with the performance of simply querying the LLM to fix the bug. As shown, the performance of \name rapidly improves and ultimately becomes comparable to the performance of \textsc{LLM-Base}, suggesting that \name shows better performance when using stronger language models; for smaller models such as CodeGen-6B, repair itself fails in a zero-shot setting, as in our experiments it would simply return the original buggy code. (We confirm that the model implementation works by also evaluating in a few-shot setting for CodeGen-6B; it could fix 44 bugs in that case.) Thus, we may speculate that as language models improve, the performance of \name will also become stronger.

\begin{tcolorbox}[boxrule=0pt,frame hidden,sharp corners,enhanced,borderline north={1pt}{0pt}{black},borderline south={1pt}{0pt}{black},boxsep=2pt,left=2pt,right=2pt,top=2.5pt,bottom=2pt]
    \textbf{Answer to RQ3:} As the underlying language model improves, the performance of \name also increases.
\end{tcolorbox}

\subsection{RQ4: Developer Benefit}
\label{sec:rq4_res}

\begin{figure*}
    \includegraphics[width=\textwidth]{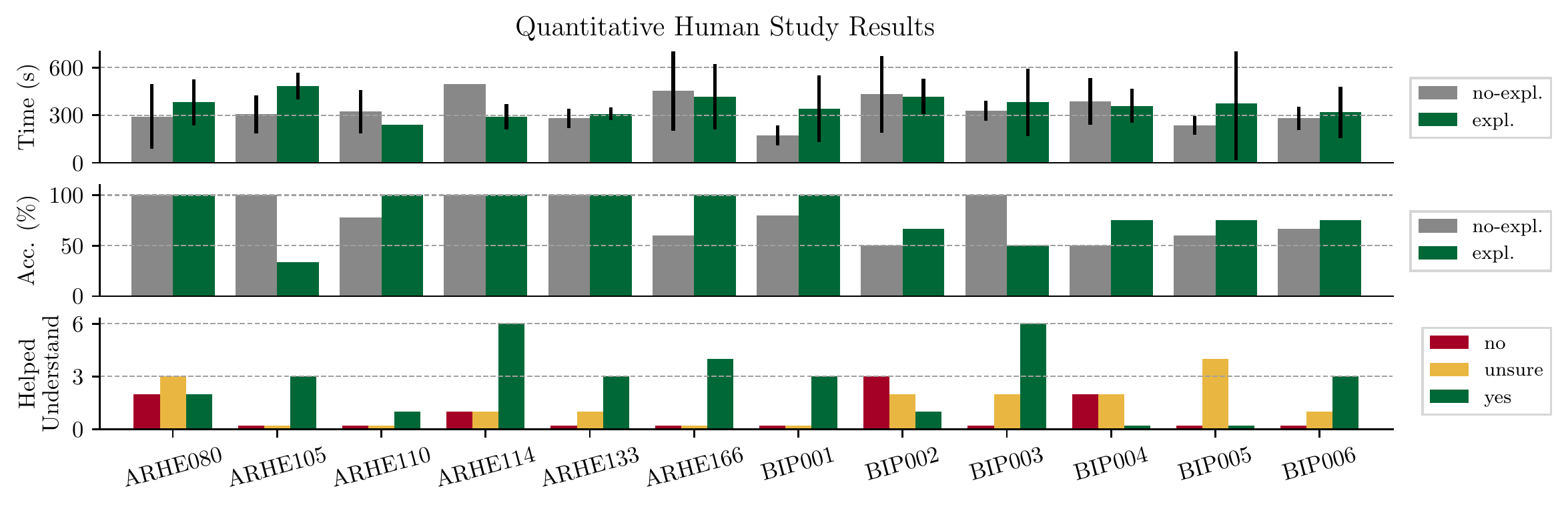}
    \caption{Developer performance on code review tasks with and without explanations from \name, and explanation ratings}
    \label{fig:hs_quantitative}
\end{figure*}

In this section, we evaluate whether developers benefit from explanations in a way that is unlikely to be swayed by a participant's opinion about explanations. The results of measuring the code review time, accuracy, and whether the explanation was rated as helpful in making the decision are presented in \Cref{fig:hs_quantitative}. 

First, looking at the amount of time that it took to solve the code review problems, we find that the time it took to solve a problem was generally similar between the case where there was no explanation and when there was an explanation. There is no case where the difference is statistically significant, despite the explanations of \name providing more information than the case without explanations, and thus potentially requiring more processing time from developers.

Regarding the accuracy with and without explanations, participants were more accurate when solving the same problems with explanations than without explanations in seven cases, with five of them being concentrated in the real-world BugsInPy benchmark. These results demonstrate that \name could have a positive impact on real-world developer productivity when using APR, as the judgment quality improved when evaluating real-world bugs while requiring roughly the same amount of developer time.
Meanwhile, there are two cases where the use of explanations lead to a drop in accuracy: ARHE105 and BIP003. For BIP003, we found that the respondents became more cautious after looking at the explanation, and answered that they needed more information to judge it. Meanwhile, for ARHE105 the participants who answered incorrectly accepted the reasoning of \name without significant scrutiny. While this was a somewhat rare incidence that happened in one of the 12 randomly sampled problems, it highlights the need of further research to identify potentially misleading reasoning. Additionally, developer accuracy improved with explanations on the two incorrect patches from BIP (BIP002 and BIP004) meaning developers are not blindly accepting patches with explanations.

On whether the participants found the explanations helpful in their decision-making, in eight of the twelve questions developers noted that the explanations were actually helpful when coming to their conclusion, underscoring the psychological benefit that providing explanations for patches holds.

\begin{tcolorbox}[boxrule=0pt,frame hidden,sharp corners,enhanced,borderline north={1pt}{0pt}{black},borderline south={1pt}{0pt}{black},boxsep=2pt,left=2pt,right=2pt,top=2.5pt,bottom=2pt]
    \textbf{Answer to RQ4:} When exposed to explanations generated by \name, human participants could process patches in roughly the same time, while achieving a higher accuracy in five of the six of the real-world bugs. They also rate the explanations as helpful in two-thirds of all bugs.
\end{tcolorbox}

\subsection{RQ5: Developer Acceptance}
\label{sec:rq5_res}

\begin{figure}[h!]
    \centering
    \begin{subfigure}[t]{0.98\linewidth}
        \centering
        \includegraphics[width=\linewidth]{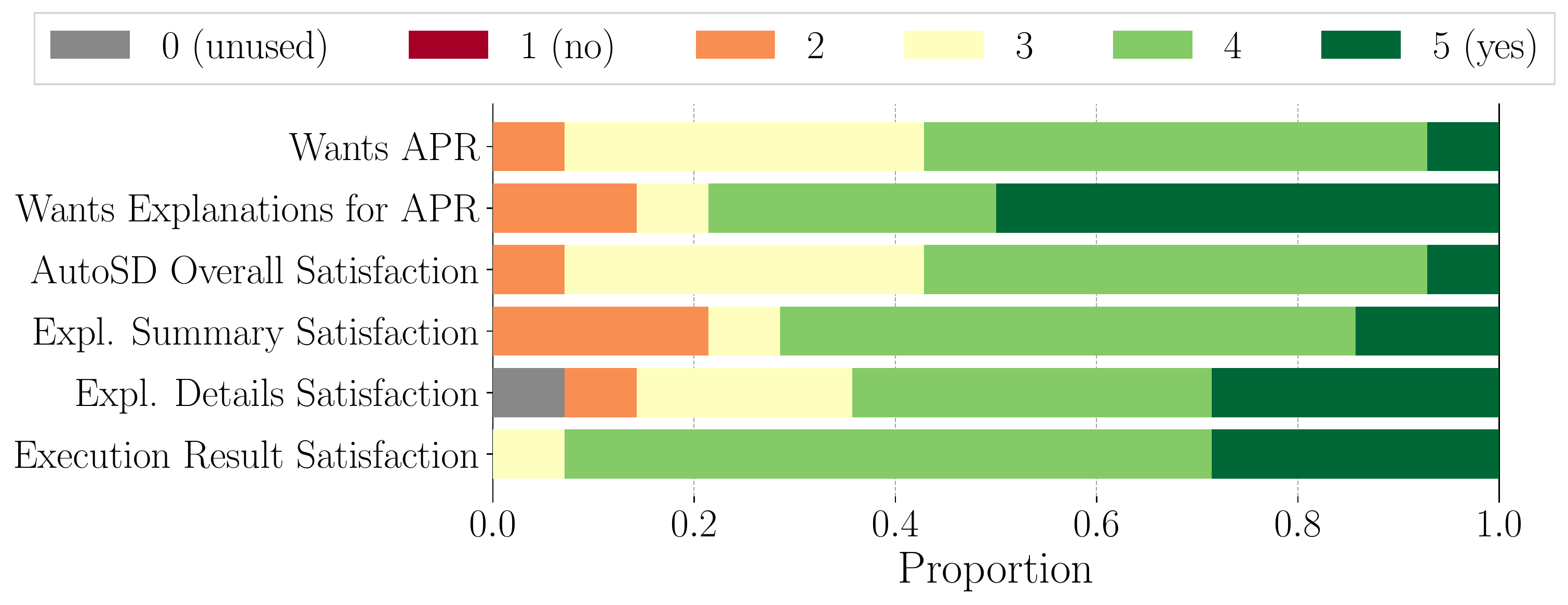}
        \caption{Satisfaction among students.}
    \end{subfigure}
    \begin{subfigure}[t]{0.95\linewidth}
        \centering
        \includegraphics[width=\linewidth]{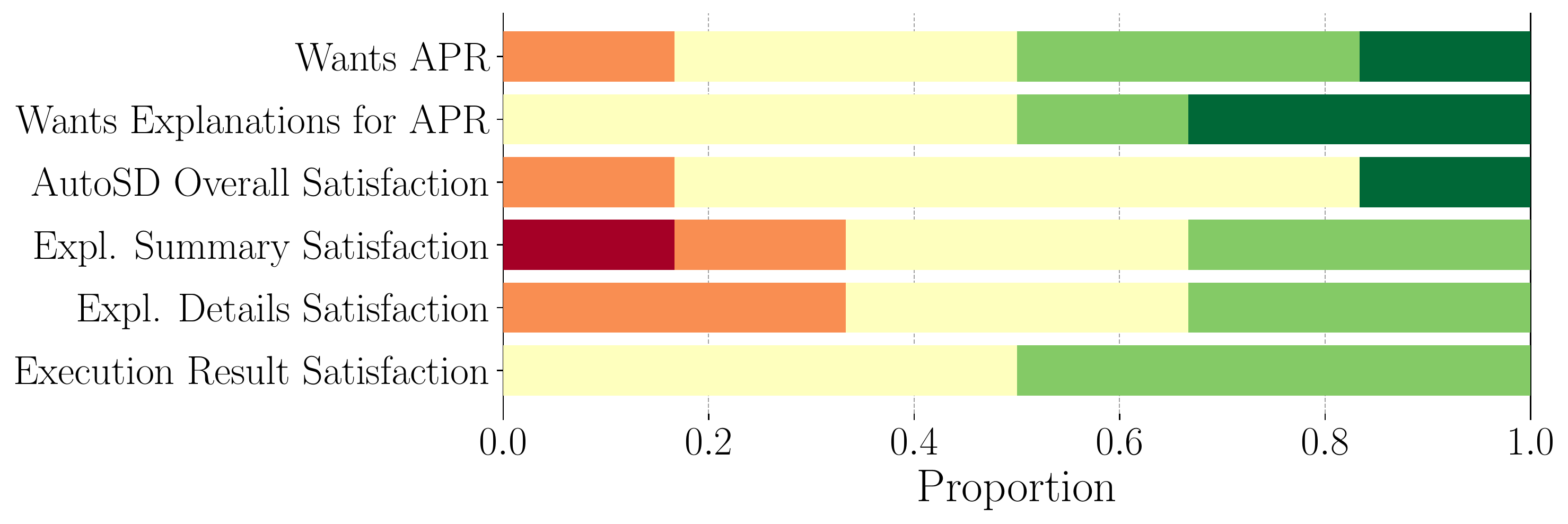}
        \caption{Satisfaction among professional developers.}
    \end{subfigure}
    \caption{Human study post-questionnaire results by group.}
    \label{fig:postq_results}
\end{figure}

The results of our post-questionnaire are presented in \Cref{fig:postq_results}. To our surprise, there was a discrepancy in satisfaction of \name between students and professional developers: while more than half of the students were satisfied with \name, only one of the six developers were satisfied. We use these differing results as an opportunity to discuss the strengths and potential improvements of \name-generated explanations.

What did students find appealing about the explanations of \name? Ten out of the 14 student participants noted that they `missed' the explanations when they were not available. When asked why they wanted to see the explanations in these cases, and how they used explanations when they were available, students described a wide range of thought processes that were aided by the existence of explanations. One common pattern was to think through the patch by oneself, then comparing one's internal thoughts to the provided explanation; one participant referred to the explanation as useful because it could function as a `rubber duck'.\footnote{See \url{https://en.wikipedia.org/wiki/Rubber_duck_debugging}.} Another common usage of explanations was to look at the explanation to discern where to focus effort on, and thus guide the direction of judgment. Other students would use the explanation to gain a better understanding of what the code was intended to do. We thus argue that a strength of \name-generated explanations is that \textbf{they can accommodate a diverse set of thought processes}, potentially aiding a wide range of developers.

Meanwhile, another usage pattern was to look at the experiments and observations within the explanations to get a concrete idea of what the values are at certain points, and use those values to build a mental model of how the bug happened. This points to another strength of \name, which is that it \textbf{incorporates actual values in its explanations}: in \Cref{fig:postq_results} (a), we note that more than 90\% of students thought that the addition of execution results improved their trust in the explanations.

On the other hand, professional developers showed a more mixed attitude towards the explanations of \name. It is noteworthy that developers are not opposed to explanations themselves: half agreed or strongly agreed that explanations would be important when using an APR tool (\Cref{fig:postq_results} (b)), highlighting the importance of the problem. When asked why they found the explanations of \name left more to be desired, one suggestion was that the current explanations would be more useful if they were connected with ``business logic'' or specifications, a suggestion echoed by one of the student participants as well. The professional developers argued that without such connections, the explanations needed to be verified rigorously and even after that were of limited value. Thus one potential direction of improvement would be to \textbf{integrate explanations with existing development artifacts like specification documents}.

Another common suggestion was to improve the interface of the tool: developers noted that they might use the tool if it was attached to an IDE, and that the explanations were too wordy. This feedback suggests that to improve developer satisfaction, we may consider \textbf{integrating explanations to platforms that developers frequent} (as also suggested by Kochhar et al.~\cite{Kochhar2016FLExpect}), and further study the specifics of explanations that developers find satisfactory.

Looking at the overall statistics, we find that 70\% of participants agreed that explanations were an important factor when using program repair, and 55\% found the scientific debugging details (Expl. Details Satisfaction of \Cref{fig:postq_results}) satisfactory, showing that a majority of participants agreed with the overall motivation and formulation of \name.

\begin{tcolorbox}[boxrule=0pt,frame hidden,sharp corners,enhanced,borderline north={1pt}{0pt}{black},borderline south={1pt}{0pt}{black},boxsep=2pt,left=2pt,right=2pt,top=2.5pt,bottom=2pt]
    \textbf{Answer to RQ5:} While the explanations of \name are capable of accommodating diverse thought processes and improving developer trust by using concrete execution results, they could be further improved by enhancing the interface and by linking to specifications.
\end{tcolorbox}

\subsection{RQ6: Qualitative Analysis}
\label{sec:rq6_res}

\begin{figure*}
    \includegraphics[width=\textwidth]{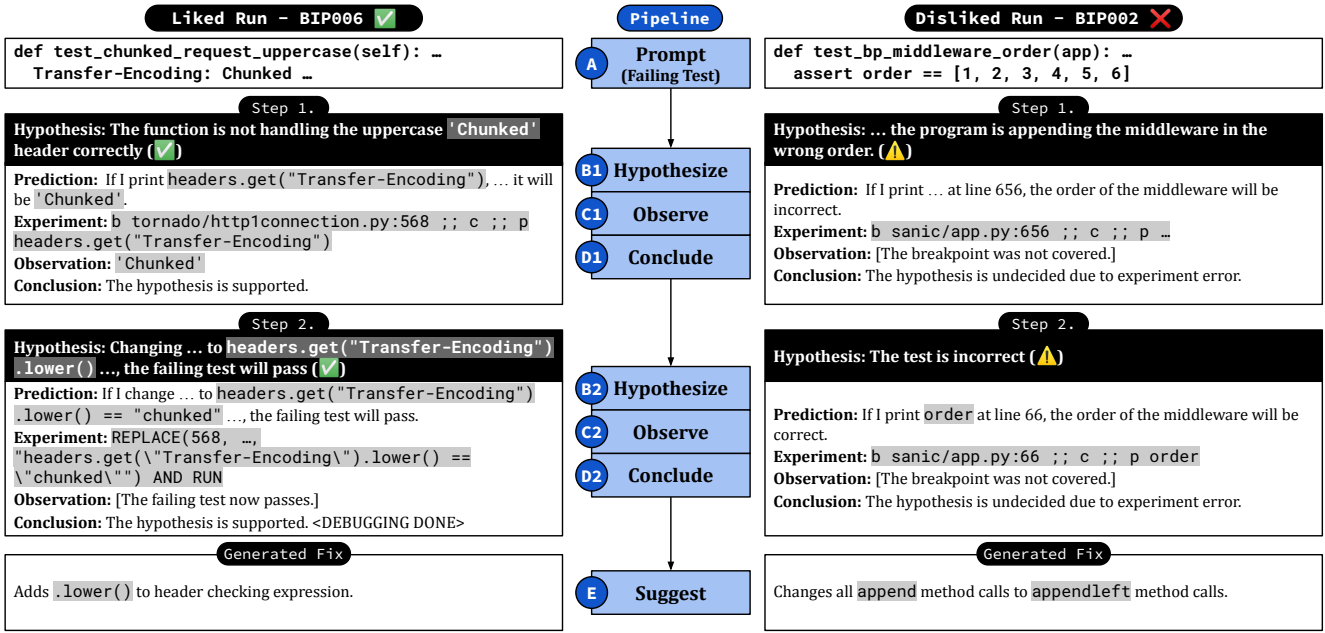}
    \caption{Example successful and unsuccessful repairs and explanations of \name from the human study.}
    \label{fig:examples}
\end{figure*}

What do the explanations generated by \name look like? In addition to the example embedded in \Cref{fig:teaser}, we provide two reasoning traces generated by \name that were liked (BIP006 - 75\% liked) and disliked (BIP002 - 16\% liked) in the human study from the real-world BugsInPy problems. On the left of \Cref{fig:examples}, we show a liked explanation, along with a condensed failing test and the generated fix. Looking at the patch, the developer will see that a \texttt{.lower()} call was added; without an explanation, this fix can appear spurious. In contrast, by providing a rationale on why \name focused on this area, participants could swiftly identify whether this fix was related to the test. For example, Student-6 said ``I first looked at the explanation, which helped me identify which part of the code to look at''. The subsequent experiment confirms that an uppercase \texttt{`Chunked'} header within the program state, which is the source of the bug. These execution results helped participants understand the bugs, e.g. Student-11 who noted that ``expression values were useful in making decisions''. Overall, this patch was correct, and the explanation aided developer comprehension and built trust. While we provide a simple example from the human study, we also note that \name works on more complex bugs as demonstrated in \Cref{sec:rq1_res}, and provide additional examples in the appendix.

Attempts at hypothesizing can fail as well. The right side of \Cref{fig:examples} depicts an case where \name fails to validate any hypotheses. While \name initially generates a hypothesis about appending in the wrong order, the line that is suggested in the experiment is actually not covered; as a result, the debugger provides feedback that the breakpoint was not covered. This is one of the most common failure causes - our analysis on 25 cases where all hypotheses were rejected revealed that in 13 of the 25 cases, breakpoints suggested by \name were never hit, and consequently \name could not get results for generated experiments. In BIP002's case, instead of looking for new breakpoints that could be covered by the test, the LLM starts suggesting that the test is wrong. Ultimately, while a fix is generated, the explanation has little connection to the patch, and as a result the human study participants rated the explanation as unhelpful; the patch itself is plausible but incorrect as well. Nonetheless, the example also illustrates how bad explanations can still lead to better decision-making: developers may see that the foundations of the patch are weak, and be (rightly) more suspicious about the patch. In this context, it is noteworthy that developers who saw the explanation of BIP002 more accurately assessed it (\Cref{fig:hs_quantitative}). Other failure modes include generating an invalid experiment expression (2/25) or adding multiple print commands in the experiment script when the infrastructure of \name only allows one print command, causing inaccurate hypothesis rejection (2/25).

\begin{tcolorbox}[boxrule=0pt,frame hidden,sharp corners,enhanced,borderline north={1pt}{0pt}{black},borderline south={1pt}{0pt}{black},boxsep=2pt,left=2pt,right=2pt,top=2.5pt,bottom=2pt]
    \textbf{Answer to RQ6:} \name can generate helpful explanations on its patches, but the reasoning process may fail as well. A common failure cause is an inability to identify the right breakpoints.
\end{tcolorbox}

\section{Discussion}
\label{sec:discussion}

This section provides threats and limitations of our work.

\subsection{Threats to Validity}
\label{sec:ttv}

\textbf{Internal Validity} concerns whether the analysis supports claims about cause and effect. Potential threats include incorrect implementations, inaccurate patch correctness assessment, and the risk of biased responses in our human study. To mitigate the impact of the first two concerns, we plan to make our implementation and repair results publicly available for scrutiny. For our human study, in addition to gathering developer sentiment about the generated explanations (which included occasional negative feedback), we also find that participant accuracy improved in five of the six BugsInPy problems, which is a result difficult to be due to bias.

\textbf{External Validity} concerns whether the results presented in this paper may generalize to other results. A particular concern when using large language models is that their training data may include segments of the evaluation data. To mitigate this issue, we newly constructed the ARHE dataset for repair and evaluated \name on that benchmark. Furthermore, our explanations were likely never within the training data, as developers usually describe code with less of a structure than Scientific Debugging prescribes, even if they think along the lines of it.

\subsection{Limitations}
\label{sec:limitations}

\name has a number of limitations that we would like to highlight. First, to enable multi-step interaction with code, both the language model and debugger must be invoked multiple times, which increases the repair time of the technique; in our experiments, \name could take about five times longer to generate a patch when compared to \textsc{LLM-Base}. Nonetheless, given the significant developer demand for explanations of automatically generated patches as shown in \Cref{fig:postq_results}, we believe that the additional cost needed to build explanations for patches is justified. Second, as a step towards explainable automatic debugging, we evaluated in the setting where method-level FL was done, and \name would then perform statement-level FL in an explainable manner. Our main focus in this paper was to establish that \name can generate explanations that aid developers in practice; we hope to work on explainable method-level FL in future work. On a related note, our technique can only handle single-method bugs as of now; incorporating a wider range of information to handle more complex bugs is also an interesting research direction. 
Finally, the generated explanation may occasionally lend credibility to incorrect patches; by allowing our technique to indicate its confidence in its output and demonstrating that confidence is correlated with correctness, we take the first steps to address this issue. Furthermore, our explanation includes concrete code execution results, aiding developer decision-making (\Cref{fig:postq_results}).

\section{Conclusion}
\label{sec:conclusion}

In this paper, we summarize the importance of explanations for automated debugging results as revealed by prior studies, and the lack of automated techniques capable of providing adequate explanations for humans. We argue this is due to a lack of automated debugging techniques that deduce in a human way, and bridge this gap between automatic and manual debugging practices by using LLMs to emulate the Scientific Debugging process. We demonstrate that \name is capable of achieving competitive repair performance when compared to other repair baselines, while having favorable properties for practical use such as an indication of confidence in the output. The repair performance of \name also improves as language models become more capable, suggesting the performance and availability of explanations may improve as language models get better. Finally, our human study reveals that the automatically generated explanations could improve developer assessment of patches, with a majority of students also expressing that they `missed' the explanations when they were not available. The interviews we performed show that the explanations \name generates could aid a wide range of developer thought patterns, and that they could be improved via tighter integration into the development process, such as making connections to written specification. Overall, we believe that the rapid improvement in language model capabilities can be harnessed to significantly ease developer use of automated techniques, and we hope to develop more human-friendly automated debugging techniques as future work.

\bibliographystyle{ACM-Reference-Format}
\bibliography{acmart}

\end{document}

\title{Appendix for Explainable Automated Debugging via\\Large Language Model-driven Scientific Debugging}

\renewcommand{\shortauthors}{Kang et al.}

\maketitle

\section{Discussion of Explainable Fault Localization}

In this section, we make the argument that fault localization needs explanations as well, and that while there are a number of fault localization techniques that argue that they are explainable and can be helpful to developers, there are few human studies. To start, as mentioned in the paper, Kochhar et al.~\cite{Kochhar2016FLExpect} survey that 85\% of developers want explanations for fault localization. If anything, the need for explanations in fault localization is greater, as while for automated program repair a suggested patch is actionable (one may accept it, reject it, or inspect it) it can be unclear what to do with a fault localization result, similarly to what has been noted for defect prediction results~\cite{Lewis2013GoogleDP}. It is noteworthy that in Kochhar et al.'s survey, developers also relate the need of an explanation to fixing and `actionability': one developer notes that ``...to make a decisions about bug fixing I want to *exactly* know why the automated tool thinks that the code have a bug [sic]'', for instance.

Some commonly used fault localization techniques include Spectrum-Based Fault Localization (SBFL) and Information Retrieval-based Fault Localization (IRFL). SBFL analyzes the coverage patterns of failing and passing tests, and uses various formulae to suggest the statements that are most correlated with the fault~\cite{Jones2002Tarantula}. Meanwhile, IRFL uses bug reports or failing tests and analyzes the textual similarity between those artifacts and the source code to identify the file or method most correlated with the failure description~\cite{Koyuncu2019IfixR}. While the reasoning traces of these techniques can be presented to developers, Kochhar et al. note that ``these basic rationales are not likely to be sufficient to help practitioners'', while citing Parnin and Orso~\cite{Parnin2011AAD}, who questioned the utility of fault localization techniques via a human study. Recent improvements in fault localization techniques include Mutation-Based Fault Localization (MBFL)~\cite{MUSE2014Moon}, which mutates statements and observes the changes in test behavior to identify likely fault locations, and fault localization based on machine learning~\cite{Xia2019DeepFL}, which uses features from an assortment of FL techniques and uses machine learning to predict which locations are likely to be faulty. Similarly to our observations about automated program repair, none of these techniques deduce in a humanlike way, and as a result it is difficult to expect that presenting the reasoning trace of any of these techniques would help understanding the results of the technique (for machine learning-based fault localization, it is also unclear if a reasoning trace exists in the first place).

At the same time, our understanding is that there is still more explainable fault localization research than in program repair. We look to three surveys~\cite{alipour2012automated,perez2014survey,Wong2016FLSurvey} to identify relevant research. \texttt{explain} was proposed by Groce~\cite{Groce2006eexplain}, which compares a failing test to the maximally similar passing test to isolate where program values diverge. Early work of Zeller was also in a related direction, in which delta debugging was applied on internal program states to perform fault localization~\cite{Zeller2002IsoCause,Hodger2005TFL}. Cleve and Zeller~\cite{Hodger2005TFL}, for example, note in the paper's conclusion that developers would ``not only know that a test has failed, but also \textit{why} and \textit{where} it failed'', indicating their interest in explaining bugs and fault localization results as well. More recently, Sumner and Zhang~\cite{sumner2013comparative} use slicing to make the state replacement technique of Zeller more precise and thus more accurately explain differences. While we are inspired by this line of work, the aforementioned literature did not perform user studies on the provided explanations that we may compare the effects against. Furthermore, relative to \name there are significant discrepancies on how debugging is done: in \name, the LLM automatically generates hypotheses about what is wrong with the code and extracts values accordingly, whereas in aforementioned work the values of all variables are inspected to isolate the bug-causing change. Whyline~\cite{Ko2008Whyline} allows developers to ask `why' and `why not' questions to a debugging system and get answers; unlike \name, the focus is not on automated debugging, and human developers are still making the hypotheses.
As a post-hoc technique that explains why a location might be buggy (but being incapable of actually describing \textit{why} a tool located that particular location) Mahbub et al. propose Bugsplainer~\cite{mahbub2023explaining}, which uses Neural Machine Translation (NMT) to train a Transformer model that translates an identified location to a likely commit message.

Overall, it seems that the observation that Alipour~\cite{alipour2012automated} made more than ten years ago that ``the most suitable level of abstraction for explaining failures is unknown'' appears to still be the case; we hope our manuscript can provide some hints to the answer.

\section{ARHE benchmark mutator breakdown}

The breakdown of the ARHE benchmark by mutation used to generate each bug is presented in \Cref{tab:arhe_breakdown}. We also describe the details of each mutator here, and compare them to mutators in PIT~\cite{Coles2016Pitest}, a widely used mutation testing tool. `Integer Literal Changer' will change literal \texttt{0} constants to \texttt{1} constants, and vice versa, which shows similar behavior to the `Inline Constant Mutator' of PIT. `If Remover' will remove the then-block or else-block of an if statement; if it has no remaining children, the if statement itself will be removed, similarly to `Remove Conditionals Mutator' of PIT. `String Literal Changer' will make a string literal empty, lower-case, or upper-case; making the string literal an empty string is not reversible, but whether the lower-casing or upper-casing can be applied in the reverse to get the original code differs from problem to problem. The generation of empty strings is similar to the `Empty returns Mutator' of PIT. `Operator Changer' will change pluses to minuses, along with similar operations, similarly to the `Math Mutator' of PIT. `Binary Operator Remover' will remove a binary operator and only leave one of the operands, similarly to the `Arithmetic Operator Deletion Mutator' of PIT. `Augmented Assignment Changer' will change \texttt{+=} to \texttt{-=}, vice versa, etc., similarly to the `Increments Mutator' of PIT. `If negator' will add a \texttt{not} to an \texttt{if} condition, similarly to the `Negate Conditionals Mutator' of PIT.

In \Cref{tab:arhe_breakdown}, the 24 bugs from If Remover and 24 bugs from Binary Operator Remover are not reversible; furthermore, we manually determine that 42 of the 63 String Literal Changer bugs are not reversible, making for a total of 90 bugs that cannot be repaired by applying the same mutation set.

\begin{table}[ht]
    \scalebox{0.9}{
    \begin{tabular}{ll}
    \toprule
    \textbf{Mutator} & \textbf{Number} \\ \midrule
    Integer Literal Changer $\circ$  & 45 \\ 
    If Remover $\square$  & 24 \\ 
    String Literal Changer $\Delta$  & 63 \\ 
    Operator Changer $\circ$ & 40 \\
    Binary Operator Remover $\square$ & 24 \\
    Augmented Assignment Changer $\circ$ & 3 \\
    If Negator $\circ$ & 1 \\
    \bottomrule
    \end{tabular}
    }
    \caption{ARHE benchmark breakdown. Reversible mutators are marked with $\circ$, unreversible mutators are marked with $\square$, and occasionally reversible mutators are marked with $\Delta$.~\label{tab:arhe_breakdown}}
\end{table}

\section{Screenshot of Website}

\begin{figure*}[h!]
    \includegraphics[width=\textwidth]{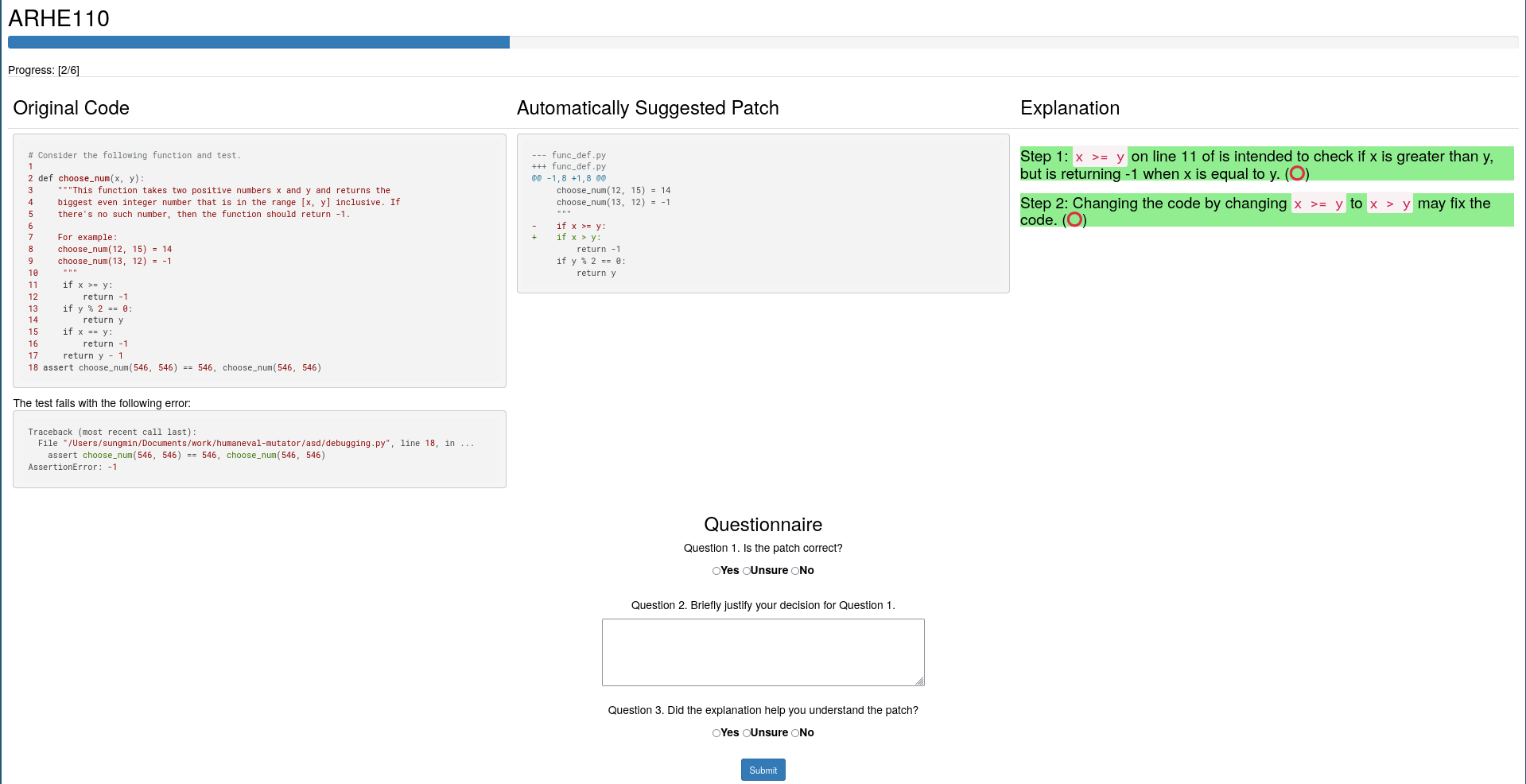}
    \caption{Screenshot of the human study webpage.}
    \label{fig:examples}
\end{figure*}

A screenshot of the human study screen is provided in \Cref{fig:examples}. Note that the original buggy code and error message are shown on the left column, the patch is suggested in the middle, the explanation is shown on the right, and the questions are presented on the bottom center of the webpage, similarly to Figure 2 of our manuscript.

\section{Scientific Debugging Prompt}

The box below shows the the Scientific Debugging description prompt used for the Defects4J benchmark.

\begin{tcolorbox}[boxrule=1pt,frame hidden,sharp corners,enhanced,borderline north={1pt}{0pt}{black},borderline south={1pt}{0pt}{black},borderline east={1pt}{0pt}{black},borderline west={1pt}{0pt}{black},boxsep=2pt,left=2pt,right=2pt,top=2.5pt,bottom=2pt,breakable]

I am going to use the scientific method to debug the problem below (as written by Zeller, 2009) by describing the hypothesis/prediction/experiment/observation/conclusion. This can be done by:

Hypothesis: An explanation for the buggy behavior. Hypotheses are the key aspect of the approach, and should be detailed and written with care. Hypotheses should build upon all previous information; repeating previous hypotheses is thus strongly discouraged. Some examples are provided below.

 - Example hypothesis 1: "Given that [information], the method is [overall erroneous behavior]. Specifically, I think it is because `c>b` on line 4321 of method `foo` is intended to [desired behavior], but is [erroneous behavior]."
 
 - Example hypothesis 2: "As the previous hypothesis was rejected, we now know `c>b` on line 4321 of the method `foo` is likely not the culprit. Looking elsewhere, perhaps `x.append(y)` should do [desired behavior], but is doing [erroneous behavior]."
 
 - Example hypothesis 3: "Because the previous hypothesis was supported, I think changing the code by changing `c>b` to `c>b \&\& a <= d` may fix the code."
 
 - Example hypothesis 4: "It seems the previous experiment ended in an error, we may need to try a different experiment. Perhaps the experiment can be refined by [new experiment]."

Prediction: A specific value or symptom that would be observed if the hypothesis is correct. Depending on the hypothesis, one may make the prediction that a test will pass. Make specific considerations for expressions within loops.

 - Example prediction 1: If I use the debugger to print [expr], while given the input and its intended role indicates that its value should be [desired value], it will not be so; that is, when I stop the debugger at line lineno, `[verifying\_expr]` will be true.
 
 - Example prediction 2: If I change [expr] to [new\_expr], the test will pass.
 
 - Example prediction 3: If I change the code to utilize the new variable, the test will pass.

Experiment: A specific `jdb` script that would check whether the prediction made is true. Make sure the line points to an actual statement (not a bracket). 
 
 - Example 1: (pdb script): `stop at org.not.a.test.class.file:lineno ; run ; print [verifying\_expr]`
 
 - Example 2: (edit script, REPLACE/ADD/DEL available): `REPLACE(4321, "c>b", "c>b \&\& a <= d") AND ADD(4323, "a+=1;") AND RUN`

Observation: The output of the `jdb` script. Example: `[value]`

Conclusion: A judgement whether the hypothesis is true based on the observation. Also add <DEBUGGING DONE> when the hypothesis confirmed leads to a concrete program fix. 
 
 - Example: The hypothesis is (supported/rejected/undecided due to experiment error). [When a test passed, add <DEBUGGING DONE>.]
\end{tcolorbox}

\section{Defects4J \name Examples}

\begin{figure*}[h!]
    \includegraphics[width=\textwidth]{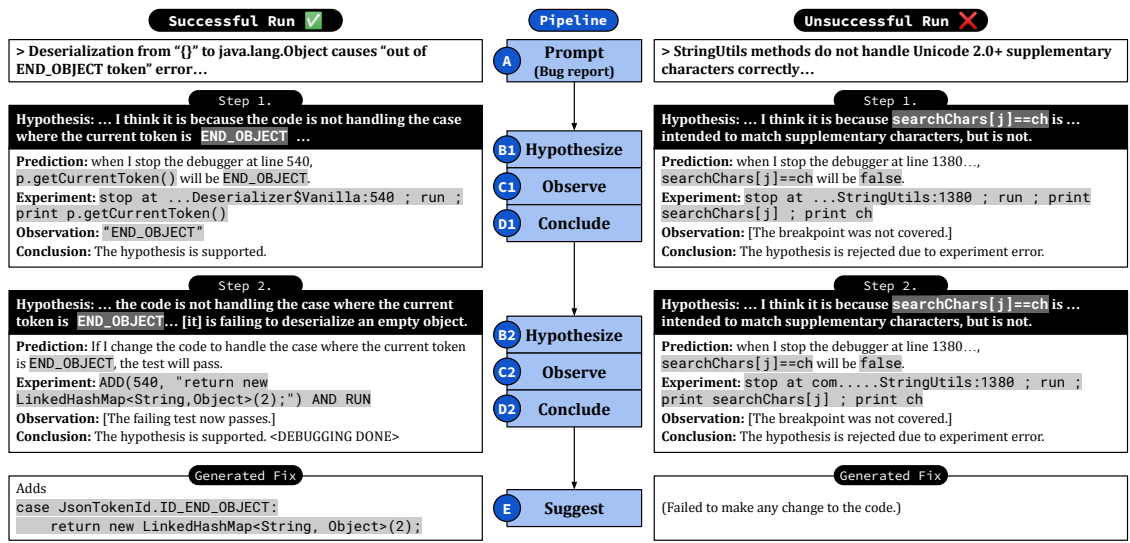}
    \caption{Example \name runs from Defects4J bugs.}
    \label{fig:d4j_examples}
\end{figure*}

A reasoning trace that ultimately lead to a correct patch and one that did not are presented in \Cref{fig:d4j_examples}. In the left case, \name hypothesizes that the bug is happening when the current token is \texttt{END\_OBJECT}, and generates an experiment to confirm that this is the case. As this is actually the case, it proceeds to search for what behavior would lead to the failing test to pass in Attempt 2. Combining these two steps together, it generates a patch identical (in this method) to the developer patch, and that makes all tests in the test suite pass. Meanwhile, on the right, another example of failing to identify the right breakpoint is provided. In this case, the same hypothesis and experiments are parroted, leading to no improvement.

\bibliographystyle{ACM-Reference-Format}
\bibliography{acmart}